# Magnetic assembly and annealing of colloidal lattices and superlattices

*Pietro Tierno**

Departament of Structure and Constituent of Matter, Universitat de Barcelona, Avenida Diagonal 647, 08028 Barcelona, Spain

Institut de Nanociència i Nanotecnologia IN$^2$UB, Universitat de Barcelona, 08028 Barcelona, Spain

E.mail: ptierno@ub.edu

**ABSTRACT**

The ability to assemble mesoscopic colloidal lattices above a surface is important for fundamental studies related with nucleation and crystallization, but also for a variety of technological applications in photonics and micro-engineering. Current techniques based on particle sedimentation above a lithographic template are limited by a slow deposition process and by the use of static templates, which make difficult to implement fast annealing procedures. Here it is demonstrated a method to realize and anneal a series of colloidal lattices displaying triangular, honeycomb or kagome-like symmetry above a structure magnetic substrate. By using a binary mixture of particles, superlattices can be realized increasing further the variety and complexity of the colloidal patterns which can be produced.

**TOC GRAPHICS**

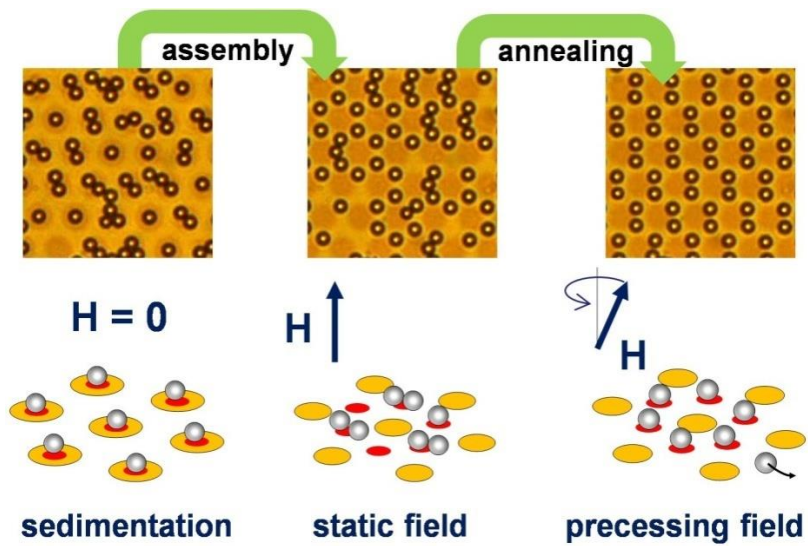

**INTRODUCTION**

Ordered two-dimensional lattices of colloidal microspheres have direct applications as photonic crystals, [1] chemical sensor, [2,3] filtration membranes, [4-6] or macroporous solids for catalysis. [7] Besides the technological interest, the ability to directly visualize crystallization process above a substrate is important to unveil the fundamental mechanisms underlying the growth of atomic and molecular thin films on a surface. [8] Highly ordered particle monolayers can be obtained via particle sedimentation above a micro-patterned template, like in colloidal epitaxy.[9] Variants of this technique have been developed in the past, based on varying the inter-particle interactions, the interactions with the substrate or the driving force which leads to the colloidal assembly.[10] However, most of these methods are usually based on the use of lithographically defined templates characterized by fixed and immovable reliefs which in principle forbid annealing the colloidal pattern via direct manipulation of the substrate topography.

This article introduces a technique to rapidly and reversibly organize two dimensional colloidal lattices with different hexagonal symmetries, e.g. triangular, honeycomb or kagome-like, above a magnetic garnet film. The garnet film, characterized by a lattice of cylindrical magnetic domains, generates a two-dimensional periodic potential which is used to attract and directly assemble magnetic microspheres. Unlike classical sedimentation of non-magnetic colloids, this method features a fast deposition process, due to the strong attraction of the magnetic substrate, and the possibility to magnetically anneal in situ the colloidal structures, reducing lattice defects and imperfections.

**EXPERIMENTAL SECTION**

To demonstrate the magnetic assembly technique, two types of monodisperse paramagnetic microspheres are used, with diameters 1.0 μm (Dynabeads Myone) and 2.8 μm (Dynabeads M-270). Both particles are composed of a highly cross-linked polystyrene matrix and are evenly doped with superparamagnetic iron-oxide grains ($\gamma$-$Fe_2O_3$ and $Fe_3O_4$). In principle, other types of magnetic particles like cubes, [11] ellipsoids, [12] Janus [13] or patchy [14] particles could be equally used.

The magnetic substrate is a ~5 μm thin single-crystal ferrite garnet film (FGF) of composition $Y_{2.5}Bi_{0.5}Fe_{5-q}Ga_qO_{12}$ (q = 0.51) and grown by standard liquid phase epitaxy on a (111) oriented gadolinium gallium garnet (GGG) substrate. [15] At equilibrium, the FGF displays a labyrinth of parallel stripe domains with magnetization vectors perpendicular to the film and periodically pointing in the upward and downward direction, spatial periodicity $\lambda$ = 6.8 micron. High frequency external fields [16] are used to transform the labyrinth pattern into a triangular lattice of cylindrical magnetic domains, also known as "magnetic bubbles", [17] with diameters *D = 6.4 μm* and lattice constant *a = 8.6 μm*. As shown in Figures 1(a,b), the magnetic bubbles are uniformly magnetized domains immersed in a film of opposite magnetization. Due to the polar Faraday effect, it is possible to visualize the size and the position of the magnetic bubbles in the FGF film via polarization microscopy.

A water drop containing a dilute solution of particles is placed on top of the FGF film, and after few minutes, the particles are two-dimensionally confined above the film due to the attraction of the magnetic domains. To avoid particle adhesion, the FGF is coated with a 1μm thick film of a UV curable photoresist (AZ-1512 Microchem, Newton, MA), via standard spin coating and UV photo-crosslink.[18] More experimental details can be found in the Supporting Information (SI).

**RESULTS AND DISCUSSION**

In absence of an external field, once deposited above the FGF, the magnetic colloids are attracted by the centers of the bubble domains, and for an area fraction $\eta = 0.09$ the particles form a triangular lattice with lattice constant $a$, Fig.1c. Here $\eta = \pi\rho d^2/4$, $d$ denotes the particle diameter, $\rho = N/A$ the two dimensional number density, $N$ the number of particles and $A = 140 \times 180 \ \mu m^2$ the area covered by the lattice. In particular, the magnetic bubble lattice used in this work was encircled by a labyrinth pattern of magnetic domains, and the FGF film has an area of $\sim 4 \ cm^2$. It is possible to extend the region of the magnetic bubbles across the whole film by improving the magnetic protocol to produce the bubble lattice. Magnetic energy calculation[19,20] shows that at the particle elevation, the FGF film presents energy minima with paraboloid-like shape located at the center of the magnetic bubbles, Fig.1f. Increasing further the particle density, each bubble becomes populated by colloidal doublets, triplets or larger compact clusters, until the particles start to form a randomly packed monolayer already for $\eta \sim 0.6$, while geometric close packing occurs at $\eta = \pi/2\sqrt{3} = 0.9$.

The magnetic bubble domains are stable even in absence of an external field, and their size can be precisely tuned upon application of a static magnetic field $H_z$ perpendicular to the film. In particular, for field strengths $H_z < 10 \ kA/m$, the diameter $D$ of the bubble domains linearly scales with the amplitude of the external field $H_z$ (Fig. S1 in SI). If $H_z$ is anti-parallel to the magnetization of the bubbles, the field decreases the size of the magnetic bubbles, increasing the area of the interstitial region. In this new configuration, the FGF allows to assemble different types of colloidal lattices, as shown in Figs. 1d and 1e. For a field $H_z = 1800$ A/m, the energy landscape of the FGF shows six regions of energy minima with triangular shape around each bubble, Fig.1g. These minima are arranged into a honeycomb lattice, i.e. a regular tiling characterized by three hexagons at each vertex. Thus for an area fraction $\eta = 0.19$, the paramagnetic colloids arrange in such a way to reproduce the underlying lattice of minima, forming a honeycomb pattern with lattice constant $a/\sqrt{3} = 5.0$ $\mu m$, Fig.1d. Increasing further the area fraction, the triangular minima start to be filled by more particles which encircle the magnetic domains filling the interstitial regions. For $\eta = 0.48$, the particles form an open lattice which resembles the kagome lattice, characterized by pairs of ordered triangles and hexagons alternating on each vertex, Fig.1e. All the colloidal lattices described until now, can be either formed via direct sedimentation in presence of the applied field, either produced by first depositing the particles, and later applying the magnetic field, as it is the case for the colloidal pattern shown in Fig.1. Thus the applied field can be used an external "knob", to control the magnetic topography of the film and, in turn, the colloidal ordering, without the need of changing the substrate, in contrast to other template-based crystallization techniques. Moreover, for applications where an immobilized lattice of particles is required, the paramagnetic colloids can be fixed above the garnet film upon addition of a small amount of salt to the water suspension, which will screen the electrostatic interaction with the surface, inducing irreversible particle sticking. The different types of lattices which can be formed by using the small (1.0 $\mu m$) particles are shown in the SI.

In order to distinguish between the various symmetries observed, the two-dimensional pair correlation function *g(r)* is calculated from the digitized particle positions, SI. Fig.2a shows the phase diagram *($\eta$,$H_z$)* encompassing all the observed structures. Beside the lattices shown in Figure 1, it was found a second triangular order for field strength within the interval *800 A/m < $H_z$ < 1600 A/*m. For these amplitudes, the energy minima inside the magnetic bubbles and in the interstitial regions have comparable depths, and a denser triangular packing can be produced for *$\eta$ = 0.3*. To the right of Figure 2 are shown as black lines the *g(r)* corresponding to the various lattices observed. Along the bottom of these graphs, are plotted as red lines the corresponding pair correlation functions for defect free lattices characterized by the same packing fraction. These correlation functions are rescaled on the y-axis so as not to interfere with the experimental ones, and display a series of sharp lines which serve as guide to classify the experimentally observed structures.

One limitation of this approach is that sedimentation alone is unable to produce a defect free crystalline structure in most of the experiments. The strong attraction of the magnetic film, with energy minima as deep as ~1100 $k_BT$ (here *T = 292 K*) makes difficult any possible particle rearrangement, whatever thermal or mechanical, e.g. via laser tweezers for example. In order to improve the colloidal ordering, a magnetic annealing procedure is developed, based on the controlled motion of particle excess within the lattice. In particular, magnetic annealing is induced via application of an external precessing magnetic field given by, ***H*** *= ($H_0$ cos($\omega t$), $H_0$ sin($\omega t$), $H_z$)*. Here $H_z$ is the amplitude of the static component perpendicular to the FGF, and $H_0$ is the amplitude of an alternating field rotating with angular frequency $\omega$ and precession angle $\vartheta$ in the *(x,y)* plane, Figure 3e. As shown in a previous work,[21] an individual particle can be propelled above a bubble lattice when subjected to an external precessing field. The field modulates the magnetic landscape in such a way that it forces the particle to pass between nearest domains at a constant speed, $V = a\omega/2\pi$. Here it is shown that, when applied to an ensemble of particles interacting via excluded volume, the precessing field is able to induce particle sorting based on the relative position above the energy minima. In particular, particles sitting above the strong triangular minima are unable to move, unless they collide with other moving colloids, while excess particles are propelled through the free pathway in the colloidal pattern, until they find empty minima to be occupied. Figs. 3(a),3(b) and Video-3.MPG in SI show a honeycomb lattice before, 3(a), and after, 3(b), the annealing procedure. Essentially, there are two mechanisms by which an excess particle can be transported towards a lattice vacancy. The first one, shown in the top row of Fig. 3c, consists in the directed motion of the excess particle, which passes from one magnetic bubble to a nearest one sliding close to the colloids forming the honeycomb lattice. The second mechanism, shown in the bottom row of Fig. 3c, consists in the transport of the excess particle via particle swapping, [20] where adjacent particles synchronously exchange their positions above a minimum, and the occupation number inside the lattice unit cell remains constant. For both mechanisms, the net particle motion ceases once the excess particle finds an unfilled minimum, or leaves the observation area. As shown in Fig. 3f, the first mechanism based on particle swapping mainly occurs when the excess particle moves along the crystallographic directions (crystallographic angles $\theta = 0° - 60°$), since along these directions are located the particles forming the honeycomb lattice. The second mechanism instead, is mainly observed at intermediate angles, where the excess particle has less probability to encounter another particle of the lattice. Fig. 3g shows that the average speed of the excess particles moving via the first mechanism can be well describe by the linear relationship $V = a\omega/2\pi$.

For the honeycomb lattice ($H_z = 1800$ A/m), the magnetic annealing via precessing field works for amplitudes of the rotating field between $800$ A/m $\leq H_0 \leq 1200$ A/m, which corresponds to precessing angles $24.0° \leq \vartheta \leq 33.7°$, and for angular frequencies $\omega < 150$ s$^{-1}$. For $\omega \geq 150$ s$^{-1}$ the particles are unable to follow the fast modulations of the magnetic landscape, and no net motion is observed. For field amplitudes $H_0 > 1200$ A/m ($\vartheta > 33.7°$), the energy minima around the magnetic bubbles merge and all particles above the FGF become mobile, inducing melting of the colloidal lattice. The process however is reversible, since the colloidal lattice can be easily and rapidly reformed just by first decreasing the amplitude of the rotating field $H_0$ and later inverting the sense of rotation of $H_0$, in order to recollect the particles which moved away from the lattice area. In contrast, for amplitudes $H_0 < 800$ A/m, the magnetic energy supplied to the system is unable to generate a net current of excess particles, and thus to anneal the colloidal lattice. The entire assembly/disassembly process is fully controlled by the external field, and the excess particles can be precisely and reversibly moved inside the lattice.

The strong attraction of the magnetic bubbles allows easily trapping and confining magnetic particles with smaller size. This would be not possible with non-magnetic colloids, since thermal forces could easily detach the particles from the substrate making more difficult the deposition process. This feature is used here to sediment particles with different sizes producing colloidal superlattices, i.e. patterns composed by different types of penetrating periodic structures.[22,23] Two illustrative examples are shown in Figure 4, with a binary mixture of paramagnetic colloids, and characterized by the area fractions $\eta_1 = 0.12$ and $\eta_2 = 0.19$, Figure 4a, and $\eta_1 = 0.02$ and $\eta_2 = 0.43$, Figure 4b. Here $\eta_1$ ($\eta_l$) denotes the area fraction of the small (large) particles. In both situations, a lattice of large magnetic colloidal particles entraps small particles at the center of the bubble domains. The small particles experience a stronger attractive force than the large ones, since are closer to the magnetic substrate, and the magnetic stray field of the bubble lattice decreases exponentially with the particle elevation. [24] In this particular case, the magnetic film generates a honeycomb lattice of minima for the large particles, and a different landscape for the small particles, characterized by energy minima at the centers of the magnetic domains. Should be noted that binary colloidal lattices have been previously reported by several groups via experiments[25,26] and numerical simulations,[27,28] in two[29, 30, 31] and three[32, 33, 34] dimensions. Although this system is limited by the magnetic film to the two-dimensional case, it allows to actuate selectively over one type of particle by varying the magnetic energy landscape. As shown in Videos-6.MPG and Videos-7.MPG in the SI, application of a precessing field with an amplitude $H_0$ = 500 A/m unable to anneal the lattice, it can be used to discriminate both types of particles, inducing only motion of the small particles in the colloidal superstructure.

**CONCLUSIONS**

In summary, a technique to engineer a series of magnetic colloidal lattices with triangular, honeycomb and kagome-like symmetry and corresponding superlattices composed of a binary particle mixture is presented. Precessing magnetic fields can be used to anneal the lattice structure either via direct transport of excess particles, either via a particle swapping mechanism, leading in both cases to a fast reduction of lattice defects and imperfections. In thin atomic and molecular films, annealing is usually induced via thermal cycling.[35] This procedure may not work with colloidal systems due to the different length-scale and interactions involved.[36] In particular, heating a colloidal crystal in order to induce defect recombination may instead provoke melting or unwanted convective currents. Should be also mention that the use of dynamic fields to anneal colloidal structures has been recently demonstrated in bulk magneto-rheological systems.[37,38] Finally, although the proposed method is demonstrated with an epitaxial growth FGF film, any other magnetic structured substrate displaying potential wells which can be tailored in deep and strength by an external field can be in principle used, if designed with the proper geometry. This includes substrates made via soft lithography, [39] electron beam lithography, [40] metal evaporation [41] or magnetron sputtering. [42]

**Figure 1.**

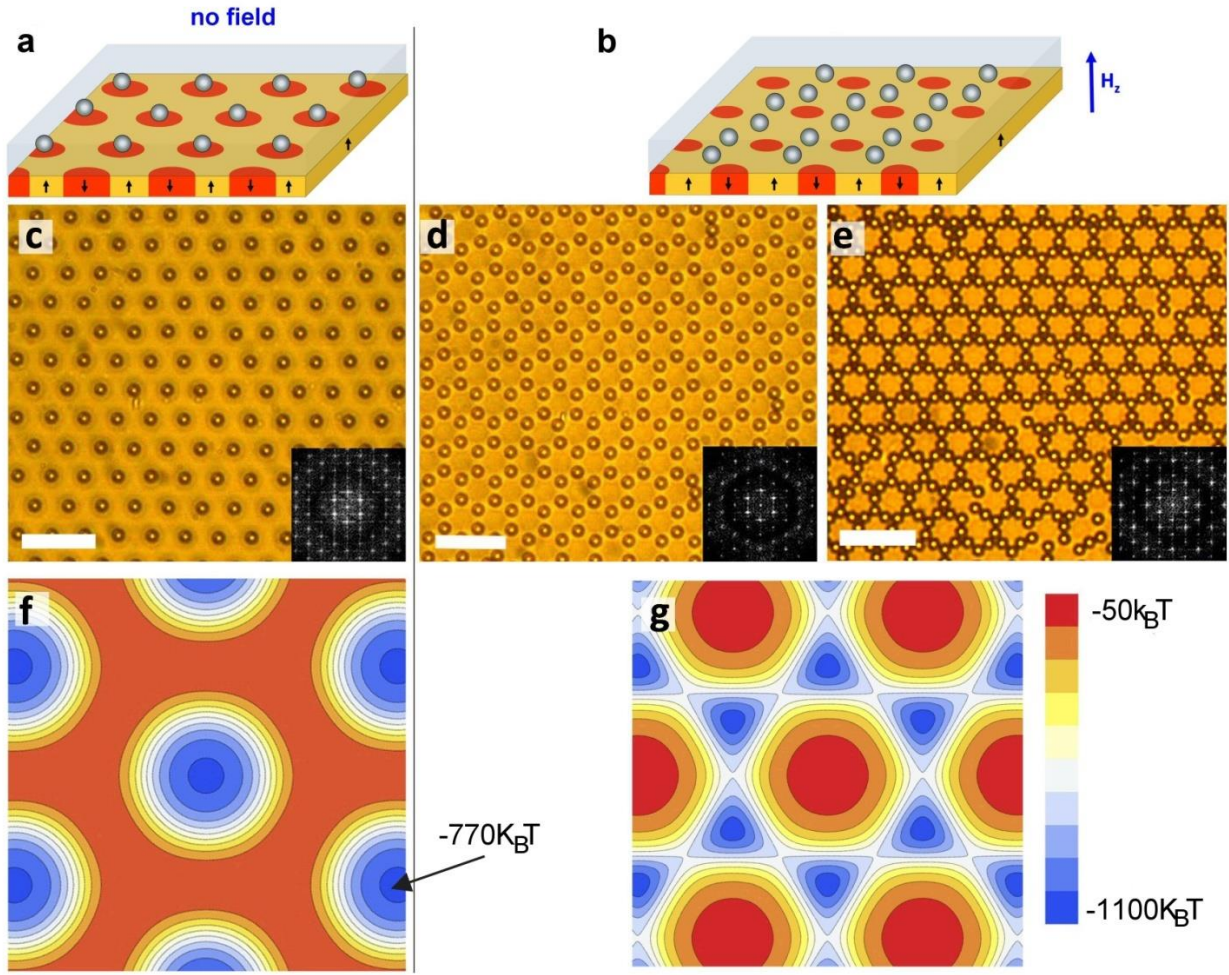

**Figure 1.** Top row: schematics showing an ensemble of paramagnetic colloid above a magnetic bubble lattice with no field (a) and with an external field $H_0$ applied perpendicular to the film (b). Middle row: triangular c ($\eta = 0.09$), honeycomb d ($\eta = 0.19$) and kagome-like e ($\eta = 0.41$) lattices composed of 2.8 µm paramagnetic colloids assembled above the magnetic bubble lattice. Lattices in Figures 1d and 1e are obtained upon application of an external field of amplitude $H_z = 1800\ A/m$. Scale bars are 20 µm, insets show FFTs of the images. Bottom row: contourplots of the magnetic potentials generated by an FGF in absence of applied field (f) and under an external field $H_z = 1800\ A/m$ (g); energy minima are in blue and maxima in red.

**Figure 2.**

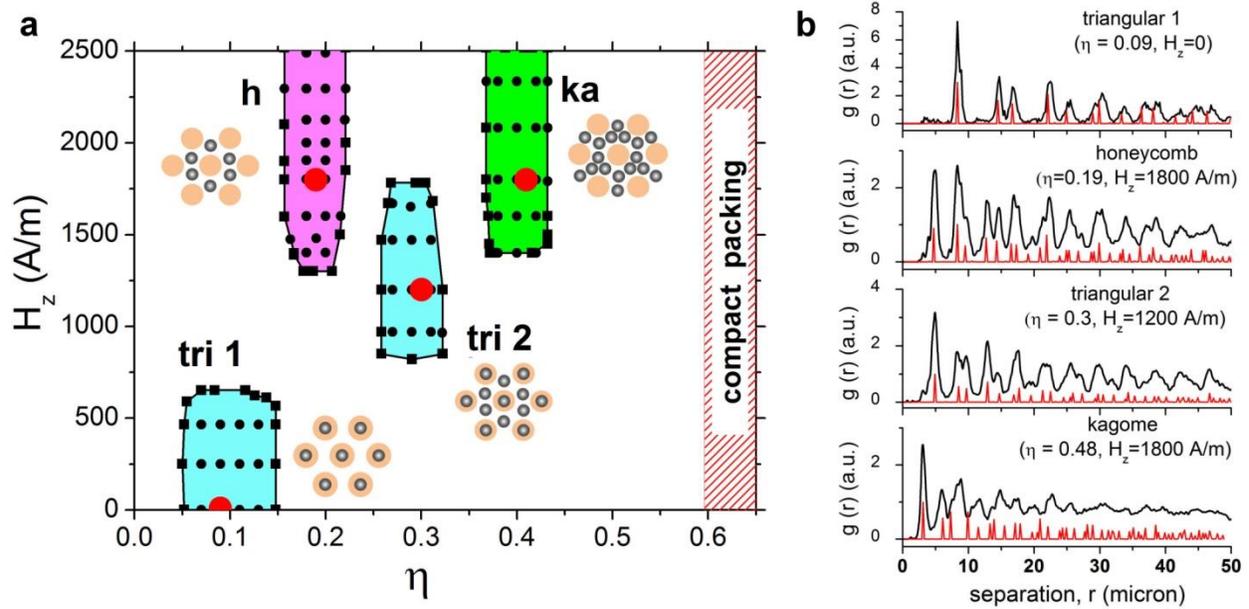

**Figure 2.**(a) Phase diagram in the *(H$_z$, η)* plane illustrating the range of parameters to observe triangular (tri), honeycomb (h) or kagome-like (ka) order. Red circles in the diagram indicate the cuts at area fractions analyzed in the graphs on the left.(b) Diagrams showing the pair correlation functions *g(r)* calculated from the particle positions (black line). Continuous red lines indicate the pair correlation functions of the corresponding defect-free lattices obtained from numerical simulations.

**Figure 3.**

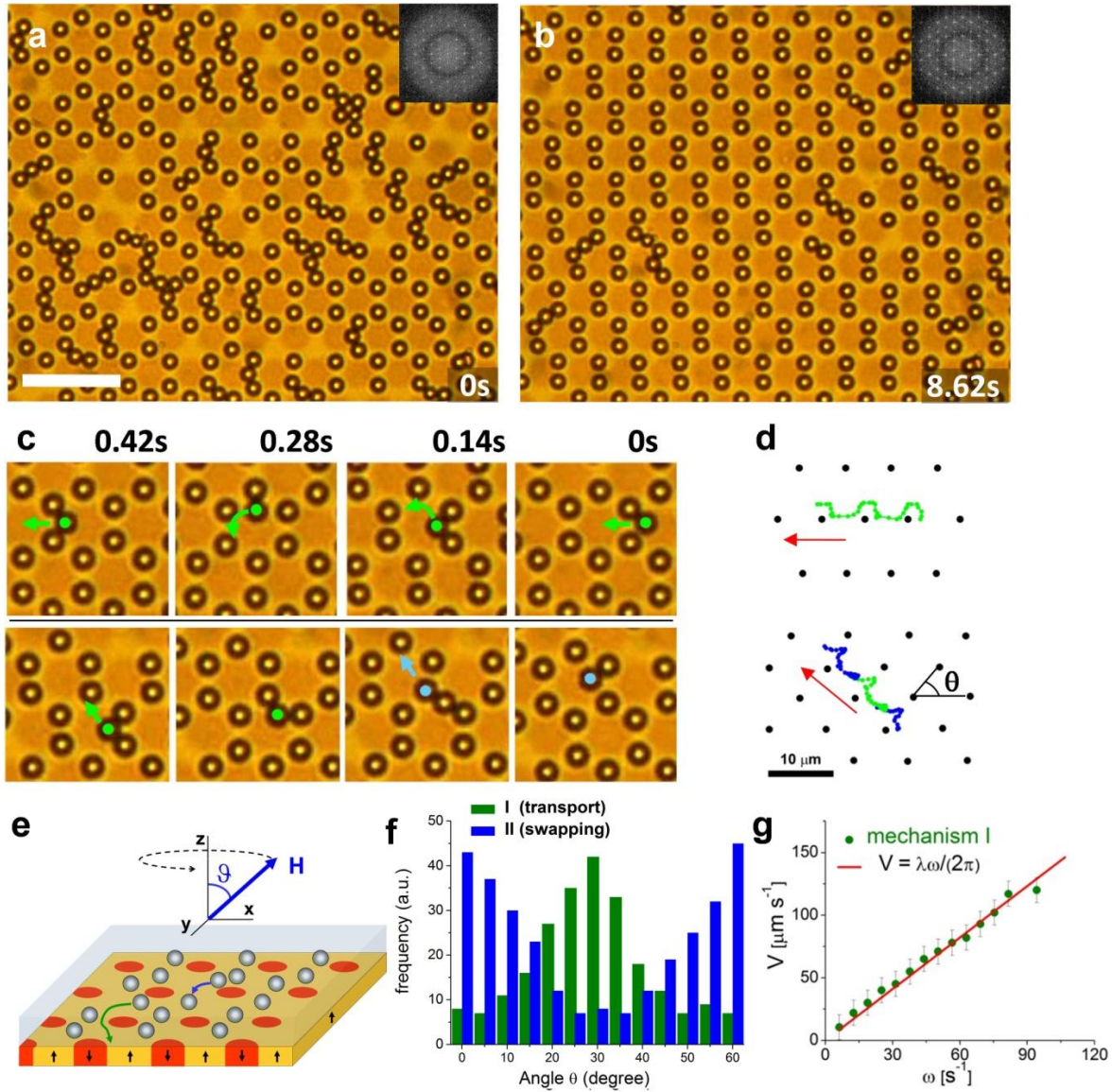

**Figure 3. a,b** Lattices of paramagnetic colloids before (a) and after (b) the annealing process based on the application of an external precessing field with frequency $\omega = 19.9\ s^{-1}$ and amplitudes $H_z = 1800\ A/m$, $H_0 = 1100\ A/m$. Scale bars is 10 μm, insets show FFTs before and after annealing. Images (c) and corresponding trajectories (d) illustrating the two types of mechanisms leading to the transport of excess particles. In both cases the time increases from right to left, different rows correspond to different mechanisms. Corresponding videos are in the SI. (e) Schematic showing the FGF film under a precessing magnetic field. (f) Histogram showing the occurrence frequency of one type of defect motion with the orientation angle $\theta$. (g) Velocity $V$ (μm s$^{-1}$) vs. driving frequency $\omega$ (s$^{-1}$) for the excess particles driven above the FGF film via the first mechanism. Continuous red line denotes $V = a\omega/2\pi$, with $a = 8.6$ μm.

**Figure 4.**

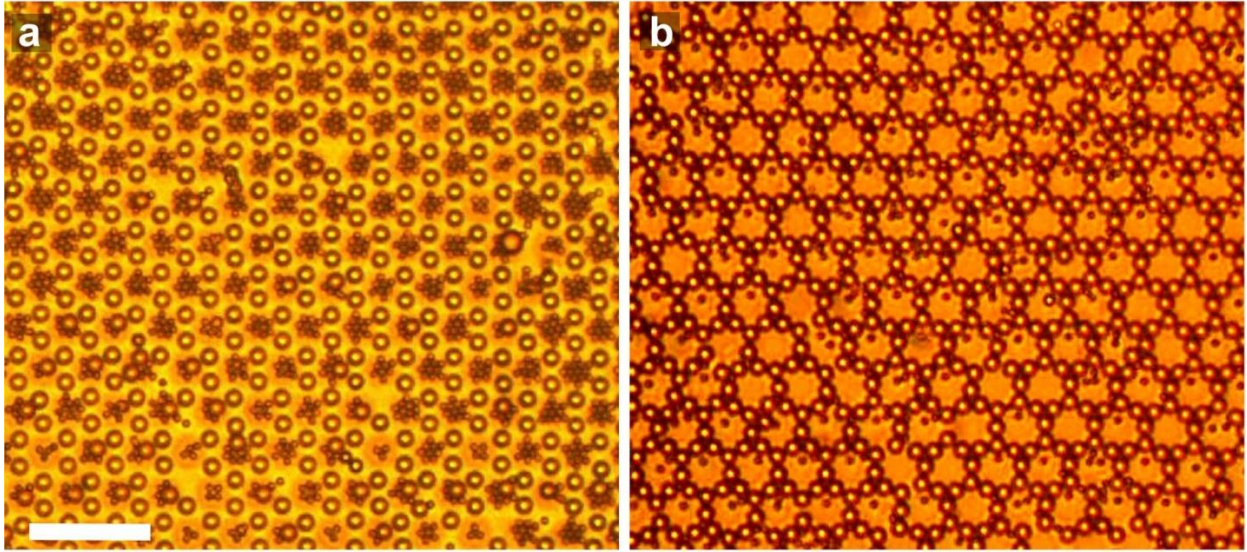

**Figure 4.** Colloidal superlattices realized by using a binary mixture of magnetic particles with 1 μm and 2.8 μm diameters. In (a) the large particles form a honeycomb lattice, in (b) a kagome-like lattice. Both superlattices are obtained under a static external field of amplitude $H_z$ = 1800 A/m. Scale bar is 20 μm, corresponding videos are in the SI.

# SUPPORTING INFORMATION

Experimental details, and data, supplementary figures. Seven movies (.MPG). This material is available free of charge via the Internet at http://pubs.acs.org.

# AUTHOR INFORMATION

**Corresponding Author**

*Email: ptierno@ub.edu

# ACKNOWLEDGMENT

The author thanks Tom H. Johansen for the FGF film. This work was supported by the ERC starting grant "DynaMO" (No. 335040), and by the MEC via programs No. RYC-2011-07605, and No. FIS2011-15948-E.

# REFERENCES

(1) Vlasov, Y. A.; Bo, X.-Z.; Sturm J. C.; Norris, D. J. On-chip natural assembly of silicon photonic bandgap crystals. *Nature* **2001**, 414, 289.

(2) Holtz, J. H.; Asher, S. A. Polymerized colloidal crystal hydrogel films as intelligent chemical sensing materials *Nature* **1997**, 389, 829

(3) Jiang, P.; Hwang, K. S.; Mittleman D. M.; Bertone, J. F.; Colvin, V. L. Template-Directed Preparation of Macroporous Polymers with Oriented and Crystalline Arrays of Voids. *J. Am. Chem. Soc.* **1999**, 121, 11630.

(4) Velev, O. D.; Jede, T. A.; Lobo, R. F.; Lenhoff, A. M. Porous silica via colloidal crystallization. *Nature* **1997**, 389, 447.

(5) Velev, O. D.; Kaler, E. W. Structured porous materials via colloidal crystal templating: From inorganic oxides to metals. *Adv. Mater.* **2000**, 12, 531.

(6) Kiesow, I.; Marczewski, D.; Reinhardt, L.; Mühlmann, M.; Possiwan, M.; Goedel, W. A. Bicontinuous Zeolite Polymer Composite Membranes Prepared via Float Casting . *J. Am. Chem. Soc.* **2013**, 135, 4380.

(7) Yu, J-S.; Kang, S.; Yoon, S. B.; Chai, G. Fabrication of Ordered Uniform Porous Carbon Networks and Their Application to a Catalyst Supporter. *J. Am. Chem. Soc.* **2004**, 124, 9382.

(8) Ganapathy, R.; Buckley, M. R.; Gerbode, S. J.; Cohen, I. Direct Measurements of Island Growth and Step-Edge Barriers in Colloidal Epitaxy. *Science* **2010**, 22, 445.

(9) Van Blaaderen, A.; Ruel, R.; Wiltzius, P. Template-directed colloidal crystallization. *Nature* **1997**, 385, 321.

(10) Li, F.; Josephson, D. P.; Stein, A.; Colloidal assembly: The road from particles to colloidal molecules and crystals. *Angew. Chem. Int. Ed.* **2011**, 50, 360.

(11) Sugimoto, T.; Khan, M.; Muramatsu, A. Preparation of monodisperse peanut-type α-$Fe_2O_3$ particles from condensed ferric hydroxide gel. *Colloids Surf. A* **1993**, 70, 167.

(12) Güell, O.; Sagués, F.; Tierno, P. Driven Janus Micro-Ellipsoids Realized via Asymmetric Gathering of the Magnetic Charge. *Adv. Mater.* **2011**, 23, 3674.

(13) Yan, J.; Chaudhary, K.; Bae, S. C.; Lewis, J. A.; Granick, S. Colloidal ribbons and rings from Janus magnetic rods. *Nat. Comm.* **2013**, 4, 1516.

(14) Zerrouki, D.; Baudry, J.; Pine, D.; Chaikin, P.; Bibette, J. Chiral colloidal clusters. *Nature* **2008**, 455, 380.

(15) Capper, P.; Mauk, M. *Liquid Phase Epitaxy of Electronic, Optical and Optoelectronic Materials*, John Wiley & Sons, 2007.

(16) W. Jantz, W.; Argyle, B. E.; Slonczewski, J. C. Spectrum and Lineshape of Bubble-Lattice Resonance. *IEEE Trans. Magn.* **1980**, 16, 657.

(17) Chang, H. *Magnetic-Bubble Memory Technology*, Marcel Dekker, 1978.

(18)     Tierno, P. Magnetically reconfigurable colloidal patterns arranged from arrays of self-assembled microscopic dimers. *Soft Matter* **2012**, 8, 11443.

(19)     Tierno, P.; Fischer, T. M. Excluded Volume Causes Integer and Fractional Plateaus in Colloidal Ratchet Currents *Phys. Rev. Lett.* **2014**, 112, 048302.

(20)     Tierno, P.; Johansen, T. H.; Fischer, T. M. Fast and rewritable colloidal assembly via field synchronized particle swapping *Appl. Phys. Lett.* **2014**, 104, 174102.

(21)     Tierno, P.; Johansen, T. H.; Fischer, T. M. Localized and Delocalized Motion of Colloidal Particles on a Magnetic Bubble Lattice. *Phys. Rev. Lett.* **2007**, 99, 038303.

(22)     Jones, M. R.; Macfarlane, R. J.; Lee, B.; Zhang, J.; Young, K. L.; Senesi, A. J.; Mirkin, C. A. DNA-nanoparticle superlattices formed from anisotropic building blocks *Nature Mater.* **2010**, 9, 913.

(23)     Dong, A. G.; Chen, J.; Vora, P. M.; Kikkawa, J. M.; Murray, C. B. Binary nanocrystal superlattice membranes self-assembled at the liquid–air interface. *Nature* **2010**, 466, 474.

(24)     Druyvesteyn, W. F.; Tjaden, D. L. A.; Dorleijn, J. W. F. Calculation of the stray field of a magnetic bubble,with application to some bubble problems. *Philips Res. Rep.* **1972**, 27, 7.

(25)     Bartlett, P.; Ottewill, R. H.; Pusey, P. N. Superlattice formation in binary mixtures of hard-sphere colloids. *Phys. Rev. Lett.* **1992**, 68, 3801.

(26)     Velikov, K.P.; Christova, C. G.; Dullens, R. P. A.; van Blaaderen, A. Layer-by-Layer Growth of Binary Colloidal Crystals. *Science*  **2002**, 296, 106.

(27)     Eldridge, M. D.; Madden, P. A.; Frenkel, D. A computer simulation investigation into the stability of theAB2superlattice in a binary hard sphere system. *Mol. Phys.* **1993**, 80, 987.

(28)     Assoud, L.; Messina, R.; Löwen, H. Stable crystalline lattices in two-dimensional binary mixtures of dipolar particles *Europhys. Lett.*, **2007**, 80, 48001.

(29)     Hoffmann, N.; Ebert, F.; Likos, C. N.; Löwen, H.; Maret, G. Partial Clustering in Binary Two-Dimensional Colloidal Suspensions. *Phys. Rev. Lett.* **2006**, 97, 078301

(30)     Law, A. D.; Martin, D.; Buzza, A.; Horozov, T. S. Two-Dimensional Colloidal Alloys. *Phys. Rev. Lett.* **2011**, 106, 128302.

(31)     Khalil, K. S.; Sagastegui, A.; Li Y., Tahir M. A.; Socolar J. E. S., Wiley B. J., Yellen B. B. Binary colloidal structures assembled through Ising interactions. *Nat. Comm.* **2012**, 3,794.

(32)     Leunissen, M. E.; Christova C. G.; Hynninen, A.-P.; Royall,C. P.; Campbell, A. I.; Imhof, A.; Dijkstra, M.; van Roij, R.; van Blaaderen, A. Ionic colloidal crystals of oppositely charged particles. *Nature* **2005**, 437, 235.

(33)     Baumgartl, J.; Dullens, R. P.; Dijkstra, M.; Roth, R.; Bechinger, C. Experimental Observation of Structural Crossover in Binary Mixtures of Colloidal Hard Spheres. *Phys. Rev. Lett.* **2007**, 98, 198303.

(34)     Bartlett, P.; Campbell, A. I. *Phys. Rev. Lett.* **2005**, 95, 128302.

(35)     *Defect Control in Semiconductors*, Edited by K. Sumino (1990 Elsevier)

(36)     Korda, P. T.; Grier, D. G. Annealing thin colloidal crystals with optical gradient forces. *J. Chem. Phys.* **2001**, 114, 7570.

(37)     Swan, J. W.; Vasquez, P. A.; Whitson, P. A.; Fincke, E. M.; Wakata, K.; Magnus, S. H.; Winne, F. D.; Barratt, M. R.; Agui, J. H.; Green, R. D.; Hall, N. R., Bohman, D. Y.; Bunnell, C. T.; Gast, A. P.; Furst E. M. Multi-scale kinetics of a field-directed colloidal phase transition. *Proc. Nat. Acad. Sci. USA* **2012**, 109, 16023;

(38)     Swan, J. W.; Bauer, J. L.; Liu, Y.; Furst, E. M. Directed colloidal self-assembly in toggled magnetic fields. *Soft Matter* **2014**,10, 1102.

(39)     Demirörs, A. F.; Pillai, P. P.; Kowalczyk, B.; Grzybowski, B. A. Colloidal assembly directed by virtual magnetic moulds. *Nature* **2013**, 503, 99.

(40)     Gunnarsson, K.; Roy, P. E.; Felton, S.; Pihl, J.; Svedlindh, P.; Berner, S.; Lidbaum, H.; Oscarsson, S. Programmable Motion and Separation of Single Magnetic Particles on Patterned Magnetic Surfaces. *Adv. Mater.* **2005**, 17, 1730.

(41)     Yellen, B.; Hovorka, O.; Friedman, G. Arranging matter by magnetic nanoparticle assemblers. *Proc. Natl. Acad. Sci. USA* **2005**, 102, 8860.

(42)     Ehresmann, A.; Lengemann, D.; Weis, T.; Albrecht, A; Langfahl-Klabes, J.; Göllner, F.; Engel, D. Asymmetric Magnetization Reversal of Stripe-Patterned Exchange Bias Layer Systems for Controlled Magnetic Particle Transport. *Adv. Mater.* **2011**, 23, 5568.